\documentclass[paper,12pt]{article}
\usepackage[centertags]{amsmath}
\usepackage{amsfonts} \usepackage{amssymb} \usepackage{amsthm}
\usepackage{graphicx}
\usepackage{cite}

\newcommand{\bra}{\langle}
\newcommand{\ket}{\rangle}

\def\one{{\rm 1\kern -.9mm l}}                             %

\def\beq{\begin{equation}}
\def\eeq{\end{equation}}
\def\beqa{\begin{eqnarray}}
\def\eeqa{\end{eqnarray}}
\newcommand{\eqa}{\begin{eqnarray}}
\newcommand{\ena}{\end{eqnarray}}
\newcommand{\p}{\partial}

\newcommand{\oo}{\mathcal{O}}

\newcommand\NN{{\mathbb N}}

\newcommand{\cC}{{\cal C}}

\newcommand{\bz}{{\bar{z}}}
\title{Dimensional Uplift in Conformal Field Theories}

\begin{document}

        \vspace*{-.6in} \thispagestyle{empty}
\begin{flushright}
\end{flushright}

\vspace{.2in} {\large
        \begin{center}
                %\resizebox{\textwidth}{!}{
                \bf  Dimensional Uplift in Conformal Field Theories
        \end{center}
}
\vspace{.2in}
\begin{center}
        {\bf 
          Ferdinando Gliozzi 
        } 
        \\
        \vspace{.2in}         {\it   
Dipartimento di Fisica dell'Universit\`a di Torino \\
Regge Center for Algebra, Geometry and Theoretical Physics,\\ 
via Pietro Giuria\, 1, I-10125 Torino, Italy   \\

\vspace{.2in}
E-mail: ferdinando.gliozzi@unito.it}

\end{center}
\vspace{.2in}

\begin{abstract}
The n-point functions of any Conformal Field Theory (CFT) in $d$ dimensions can
always be interpreted as spatial restrictions of corresponding functions in a
higher-dimensional CFT with dimension $d'> d$. In particular, when a four-point
function in $d$ dimensions has a known conformal block expansion, this expansion can
be easily extended to $d'=d+2$ due to a remarkable identity among conformal blocks,
 discovered by Kaviraj, Rychkov and Trevisani (KRT) as a consequence of
Parisi-Sourlas supersymmetry and  confirmed to hold in any CFT with $d > 1$.

In this note, we provide an elementary proof of this identity using simple
algebraic properties of the Casimir operators. Additionally, we construct five
differential operators,  $\Lambda_i$, which promote a conformal block in
$d$ dimensions to five conformal blocks in $d+2$ dimensions. These operators can
be normalized such that $\sum_i \Lambda_i = 1$, from which the KRT
identity immediately follows. Similar, simpler identities have been proposed,
all of which can be reformulated in the same way.

\end{abstract}
\vspace{.3in}
\hspace{0.7cm} 
%{\small \setlength{\parskip}{0pt} \tableofcontents}
\newpage

\section{Introduction}
Dimensional reduction is a property emerging in some disordered systems.  In the
random-field Ising model it was observed that the correlation functions 
 reduce to those of the pure Ising model, without the random source,
 in two fewer dimensions \cite{Grinstein:1976zz,Aharony:1976jx}. This was
understood by Parisi and Sourlas \cite{Parisi:1979ka}
to be due to  a hidden supersymmetric formulation. This dimensional
reduction holds
near the upper critical dimension of six, but
 fails in sufficiently low dimensions 
\cite{Imbrie:1984ki,Brezin:1998fb,Tissier:2011mv,Baczyk:2013uda,Cardy:1985,Fytas:2019zdk,Kaviraj:2019tbg,Kaviraj:2020pwv,Kaviraj:2021qii}. 
This failure is now
 believed to result from operators that are irrelevant near six
 dimensions becoming relevant around five dimensions \cite{Tissier:2011mv,Kaviraj:2019tbg,Wiese:2021qpk}. 

Another notable example is the  connection, again discovered
by Parisi and Sourlas \cite{Parisi:1980ia}, between randomly branched polymers 
or lattice animals in $d$ dimensions \cite{Cardy:2001ci,Cardy:2003,hsu2005,luther2011,Kaviraj:2022bvd} and the Yang-Lee edge
singularity \cite{Cardy:2023lha} in two fewer dimensions. In this
case, dimensional reduction works not only near the upper critical dimension of
eight but continues down to two dimensions, as rigorously demonstrated
by Brydges and Imbrie \cite{Brydges2003a,Brydges2003b}. Even in these
instances, the correlation functions on branched polymers reduce to
those of the Yang-Lee edge singularity.

In the present  note we explore such a connection across the space
dimensions from the point of view of conformal field theory (CFT). 
As a  result, we give a simple algebraic proof of a surprising exact
relation discovered by Kaviraj, Rychkov and Trevisani (KRT) as a
direct consequence of the Parisi-Sourlas (PS) supersymmetry and confirmed to hold in
any CFT with $d>1$ \cite{Kaviraj:2019tbg}; this relation expresses any  $(d-2)$-dimensional
conformal block  as a linear combination of
five conformal blocks in dimension $d$. Being these functions defined
as eigenfunctions of the Casimir operators, we study some algebraic
properties of these operators instead of working directly on
the conformal blocks like in the original proof of Ref.
\cite{Kaviraj:2019tbg}.  We in a sense
 reverse the KRT identity by constructing, only utilizing some
commutation properties of the Casimir operators, five
differential operators, $\Lambda_i$,  transforming an arbitrary 
$(d-2)$-dimensional conformal block $g^{(d-2)}_{\Delta,\ell}$
( describing the contribution to a four-point function of a primary of scaling dimension
$\Delta$ and spin $\ell$) into five
different conformal blocks in dimension $d$,
                 \beq\Lambda_i\,\, g^{(d-2)}_{\Delta, \ell}
\propto\, g^{(d)}_{\Delta_i,\ell_i}\,,  \eeq
where the  conformal blocks in the {\sl rhs} are precisely those 
contributing to the KRT identity. Equation (1) can be utilized to
build new conformal blocks in higher
dimensions once their form in lower dimensions is known.  
The $\Lambda_i$'s  can be normalized so that
\beq
                      \sum_{i=1}^5 \Lambda_i=1~~,\eeq
therefore $\sum_i\Lambda_i\,g^{(d-2)}_{\Delta,\ell}$ yields at once the KRT identity.  In such a form it is clear that this identity holds
for any real $d>1$ under the only assumption that the conformal blocks
are eigenfunctions of the Casimir operators.  

The KRT identity allows one to reconstruct the spectrum of primary operators
contributing to a four-point function of a CFT in dimension  $d$ once
this spectrum is known in a $(d-2)$-dimensional CFT with the same
external scalars \cite{Trevisani:2024djr}.
\footnote{In \cite{Trevisani:2024djr} this result is obtained by assuming the PS supersymmetry.}

\section{Notation }
 In a generic CFT in $d$ dimensions the 4pt function of arbitrary scalars
$\oo_i$ of scaling dimension $\Delta_i$ can be parametrised as \cite{Dolan:2011dv,Hogervorst:2013kva}
\beq
\bra \oo_1(x_1)\oo_2(x_2)\oo_3(x_3)\oo_4(x_4)\ket =K(x_i)\,g(u, v)\,,
\eeq
where $K(x_i)$ is  a kinematic factor given by
\beq
K(x_i)=\frac1{\vert x_{12}\vert^{\Delta_1+\Delta_2}\,\vert x_{34}\vert^{\Delta_3+\Delta_4}}\,\left(\frac{x_{14}^2}{x_{24}^2}\right)^a\,\left(\frac{x_{14}^2}{x_{13}^2}\right)^b\,,
\eeq
with $a=\frac{\Delta_2-\Delta_1}2$ and
$b=\frac{\Delta_3-\Delta_4}2$\,; $g(u,v)$ is a theory-dependent
function of the two cross-ratios $u=\frac{x_{12}^2 x_{34}^2}{x_{13}^2 x_{24}^2}$ and
  $v=\frac{x_{14}^2x_{23}^2}{x_{13}^2x_{24}^2} $, which are linearly independent for $d>1$. 

It is important to emphasize that although 
$g(u,v)$ depends on the scaling dimensions of the external scalars, it is completely
independent of the space-time dimension $d$:
\beq
\p_d\,\,g(u,v)=0\,\,
\eeq
In principle, we can choose $d$ to be any real number, provided that 
$d\ge\sf d$, where $\sf d$ represents the dimension of the linear space spanned
by the four points $x_i$
  (hence, $1<\sf d\le 3$). Under this condition, we obtain the expansion
\beq
g(u,v)= \sum_{\Delta,\ell}\,c_{\Delta,\,\ell}\,\,g^{(d)}_{\Delta,\ell}(u,v),
\eeq
where the coefficients $c_{\Delta,\,\ell}$ are the operator product
expansion (OPE) coefficients.

The freedom in choosing $d$ has intriguing physical consequences. For
instance,  if 
$g(u,v)$ corresponds to the four-point function of a random branched
polymer system in dimension $d$, then, due to dimensional reduction, 
its expansion in $(d-2)$-dimensional conformal blocks provides insight 
into the spectrum of primary operators associated with the Yang-Lee 
edge singularity. The dimensional shift by 2 in this case is well 
understood in terms of Parisi-Sourlas supersymmetry.

A natural question then arises: can this reduction by 2, or the
reverse approach, the dimensional uplift, be explained purely through
conformal invariance? A simple example helps to clarify the issue. 
Consider the four-point function 
$\bra \phi(x_1)\phi(x_2)\phi(x_3)\phi(x_4)\ket\,$,
where $\phi$ is a scalar generalized  free field (GFF) with an
arbitrary scaling dimension $\delta$. The corresponding function 
$g(u,v)$ is given by
\beq
g(u,v) = 1 + u^\delta + \left(\frac{u}{v}\right)^\delta.
\eeq
The OPE coefficients and the spectrum of contributing primaries have
been exactly evaluated 
in various ways \cite{Fitzpatrick:2011dm,Gliozzi:2016ysv,Gliozzi:2017hni}.
Since $\delta$ is an arbitrary real number, generalized free CFTs are not
necessarily local CFTs. For locality, the theory must have a
non-vanishing coupling to the
conserved energy-momentum tensor $T_{\mu\,\nu}$, which is a primary operator with
scaling dimension $d$ and spin 2.
The spectrum of primaries  contributing to the expansion of 
$g(u,v)$ is
\beq
\{\Delta,\ell\} = \{2\delta + 2m + 2n, \ell = 2n\}, \quad (m,n=0,1,2,\dots)\,.
\eeq

Keeping $\delta$ fixed, if the equation 
$2\delta+2m+2=d$ has a solution, then clearly there exist infinitely many solutions
for dimensions $d'=d+2k$ with $k=1,2,\dots$ .

This behavior seems to be a general feature of local conformal field
theories. Actually, as
already pointed out in \cite{Kaviraj:2019tbg}, the KRT identity
implies  that if a scalar primary of scaling
dimension $\delta$ has a non-vanishing coupling to the energy-momentum 
tensor in dimension $d$, then it describes a {\sl local} CFT in any
space dimension $d'=d+2k\,\,$ , $(k\in\NN)$\,, and this property holds
for any {\sl real} $d>1$. 

\section{The proof}
The quadratic Casimir operator $C_2[d]$ can be written as \cite{Dolan:2011dv}
\beq
C_2[d]=D_z+D_\bz+(d-2)\frac{z\,\bz}{z-\bz}\left((1-z)\p_z-(1-\bz)\p_\bz\right)\,,
\eeq
with
\beq
D_z=z^2(1-z)\p_z^2-(1+a+b)z^2\p_z+ab\,z~.
\eeq
 The two variables $z,\bz$ are related to the cross-ratios as $u=z\,\bz$
and $v=(1-z)(1-\bz)$.

We aim to demonstrate a special relationship between the quadratic Casimir operators
in $d$ and $d-2$ dimensions. Specifically, we  construct iteratively
five differential operators $Y_i$ $(i=1,\dots 5)$ where 
$Y_1=1$, $Y_2$ is first-order in $z$ and $\bz$, $Y_3$ and $Y_4$ are
second-order, and $Y_5$ is third-order, such that

\beq
C_2[d]\,Y_i=Y_i\,C_2[d-2]+Y_j\,\alpha^j_i\,\,\,,
\label{e11}
\eeq
where the $\alpha^j_i$'s are simple linear functions of the Casimir operators 
in dimension $d-2$. If we attempt to replace on the right-hand side
(RHS) $C_2[d-2]$ with  $C_2[d']$
for $d'\ne d-2$, the linearity in the $Y_i$
 is immediately lost, and additional higher-order differential
 operators are introduced.

When both sides of Equation (\ref{e11}) act on the conformal block
$g^{(d-2)}_{\Delta,\ell}$, the Casimir operators on the RHS can be replaced
by their eigenvalues. In this case, the $\alpha^j_i$'s become real or complex
numbers, and   Eq. (\ref{e11}) reduces to
\beq
C_2[d]\,Y_i\,g^{(d-2)}_{\Delta,\ell}=a^j_i\,Y_j\,
g^{(d-2)}_{\Delta,\ell}\,\,\,.
\label{e12}
\eeq
This means that $C_2[d]$ acts on the linear space spanned by the five 
functions $Y_i\,g^{(d-2)}_{\Delta,\ell}$ as the matrix $a^j_i$. In particular, the
five left-eigenvectors $V_i$ of $a^j_i$ allow us to define five
differential operators 
\beq
\Lambda_i\equiv\, V_i^j\,Y_j\,\,.
\label{e13}
\eeq
 These operators promote the conformal block $g^{(d-2)}_{\Delta,\ell}$ 
 to appropriate conformal blocks in 
$d$ dimensions, as outlined in the Introduction.

  We begin  with  $Y_1=1$ and write
\beq
C_2[d]\,Y_1=Y_1\,C_2[d-2]+Y_2~.
\label{f1}
\eeq
Here, $Y_2=\frac{2
  z\,\bz}{z-\bz}\left((1-z)\,\p_z-(1-\bz)\,\p_\bz\right)$ is a
first-order differential operator.

The action of $C_2[d]$ on $Y_2$  yields a second-order operator, 
which we write for convenience as $Y_3-Y_4$. 
Specifically, we obtain
\beq
C_2[d]\,Y_2=Y_2\,(C_2[d-2]-d+2)+2\,Y_1\, C_2[d-2]+2\, Y_3-2\, Y_4~,
\label{f2}
\eeq
where $Y_3=\frac{z\,\bz}{z-\bz}(D_z-D_\bz)$ and $Y_4=\frac{z+\bz}{z
  \bz}\, Y_3$. 

We now   have
\beq
C_2[d]\,Y_3= Y_3\,(C_2[d-2]-d+2)~;
\label{f3}
\eeq
this remarkable identity, first discovered by Dolan and Osborn \cite{Dolan:2011dv} and
recently derived by Trevisani \cite{Trevisani:2024djr} as  a consequence of the PS
supersymmetry, shows that $Y_3\,g^{(d-2)}_{\Delta,\ell}$ is
proportional to $g^{(d)}_{\Delta+1,\ell-1}$, hence  it  already
corresponds to one of the desired $\Lambda_i$.

$C_2[d]\,Y_4$ yields a third-order operator, which we denote by
$Y_5$. 
We find
\beq
C_2[d]\, Y_4= Y_4\,(C_2[d-2]+2)+Y_3\,(2(a+b)-d)+2\, Y_5~,
\label{f4}
\eeq
where $Y_5=((z-1)\,\p_z+(\bz-1)\,\p_\bz)Y_3$. 

Finally, applying $C_2[d]$ to $Y_5$ generates a fourth-order operator.
However this operator turns out to be  the
quartic Casimir operator in $(d-2)$ dimensions. Specifically we find
\eqa
C_2[d]\,Y_5=&Y_5\,(C_2[d-2]-3d+8)+Y_4\,(C_2[d-2]-(d-2)^2)-Y_1\,C_4[d-2]\nonumber\\
&+Y_3\,(2a b-2(d-3)(a+b)+(d-2)^2-C_2[d-2])~.
\label{f5}
\ena
Here, $C_4$ is the quartic Casimir operator as defined by Dolan and
Osborn in  \cite{Dolan:2011dv} :
\beq
C_4[d]=\left(\frac{z
\bz}{z-\bz}\right)^{d-2}\left(D_z-D_\bz\right)\left(\frac{z
\bz}{z-\bz}\right)^{-d+1}\,Y_3~.
\eeq
 When Eq.(\ref{f5}) is applied  to $g^{d-2}_{\Delta,\ell}$, 
$\,C_4[d-2]$ is replaced by its eigenvalue, therefore our iterative construction
of $Y_i$'s terminates at $i=5$ and no higher-order differential
operators arise. 
 
The five  Eq.s (\ref{f1}-\ref{f5}), when acting on
$g^{(d-2)}_{\Delta,\ell}$, can be recast into the
matrix form given in (\ref{e12}), namely 
\beq
C_2[d] {\sf Y}=\textstyle\left(\begin{matrix}
c_2&1&0&0&0\cr
2 c_2&c_2-d+2&2&-2&0\cr
0&0&c_2-d+2&0&0\cr
0&0&2(a+b)-d&c_2+2&2\cr
-c_4&0&\kappa&c_2-(d-2)^2&c_2-3d+8\cr
\end{matrix}
\right) {\sf Y}
\label{matrix}
\eeq
where ${\sf Y}$ is the column-vector formed by the $Y_i$'s, $\kappa$ is the 
coefficient of
$Y_3$ in Eq.(\ref{f5}) and $c_2,c_4$ are the eigenvalues of the quadratic and quartic Casimir
operators in $d-2$ dimensions \cite{Dolan:2011dv}:
\eqa
c_2=&\ell(\ell+d-4)/2+\Delta(\Delta-d+2)/2\,\,\,;\\
c_4=&-\ell(\ell+d-4)(d-\Delta-3)(\Delta-1)\,\,\,\,.
\ena

As a consistency check, one can verify that the five eigenvalues of the
matrix (\ref{matrix}) correspond to the five primary operators in $d$
dimensions contributing to the KRT identity, namely
\beq
\Delta_i=\Delta
+\alpha\,\,\,\,,\,\,\ell_i=\ell-\beta\,\,\,,\alpha,\beta\,\in\{
0,1,2\},\,\,\alpha+\beta\,\,\,{\rm even}\,.
\label{iab}
\eeq

Following Eq. (\ref{e13}) we can compute  the five left-eigenvectors
of the matrix and  define five differential operators 
$\Lambda_i=V_i\cdot Y$ which
transform $g^{(d-2)}_{\Delta,\ell}$ into the corresponding conformal block in
$d$ dimensions. It is convenient to replace the index $i$ with the two
indices $\alpha,\beta$  defined in Eq.(\ref{iab}) which describe the
action of $\Lambda_i$ on $\Delta$ and $\ell$:  
\beq
\Lambda_i\,g^{(d-2)}_{\Delta,\ell}\equiv\Lambda_{\alpha\,\beta}\,\,g^{(d-2)}_{\Delta,\ell}\propto
g^{(d)}_{\Delta+\alpha,\ell-\beta}\,,\,\alpha,\beta\in\{0,1,2\},\,
\alpha+\beta\,\,\,{\rm even}\,.
\eeq
Precisely we obtain
\begin{align}
\Lambda_{00}&=\textstyle~n_{00}\,\big[ Y_5+(3d-8+\ell-\Delta)\,Y_4+(a+b+\frac{2a\,b}{d-2+\ell-\Delta}-\frac{3d-8+\ell-\Delta}2)\,Y_3+\nonumber\\
&\textstyle~~~~~~~~~~~~~~~~~~~~~~~~~~~~~~~~ +\frac{(d-4+\ell)(3-d+\Delta)}{2}\,Y_2+\frac{(d-4+\ell)(3-d+\Delta)(d+\ell-\Delta-2)}2\big]\,, \nonumber\\
\Lambda_{20}&=\textstyle~n_{20}\big[Y_5+\frac{2d-6+\ell+\Delta}2\,Y_4+(3+a+b+\frac{2a\,b}{\ell+\Delta}-\frac{\Delta+\ell}{2})\,Y_3-\nonumber\\
&\textstyle~~~~~~~~~~~~~~~~~~~~~~~~~~~~~~~~~-\frac{(\Delta-1)(d-4+\ell)}{2}\,Y_2-\frac{(d-4+\ell)(\Delta-1)(\Delta+\ell)}{2}\big]\,,\nonumber\\
\Lambda_{11}&=\textstyle~n_{11}\,Y_3\,,\\
\Lambda_{02}&=\textstyle~n_{02}\big[Y_5+(d-2-\frac{\ell+\Delta}{2})\,Y_4+(a+b-\frac{2
a\,b}{\ell+\Delta-2}+\frac{4-2d+\ell+\Delta}{2}+\ell+\Delta-2)\,Y_3+\nonumber\\
&\textstyle~~~~~~~~~~~~~~~~~~~~~~~~~~~~~~~~~~~+\frac{\ell(d-3-\Delta)}2\,Y_2-\frac{\ell(d-3-\Delta)(\Delta+\ell-2)}2\big]\,,\nonumber\\
\Lambda_{22}&=\textstyle~n_{22}\big[Y_5+\frac{d-2-\ell+\Delta}2\,Y_4+\frac{2(a+b+1)+\ell-\Delta-d-\frac{4a\,b}{d-4+\ell-\Delta}}2\,Y_3+\nonumber\\
&\textstyle~~~~~~~~~~~~~~~~~~~~~~~~~~~~~~~~~~~~+\frac{\ell(\Delta-1)}2\,Y_2-\frac{\ell(\Delta-1)(d-4+\ell-\Delta)}2\big]\,.\nonumber
\end{align}

The normalization coefficients $n_{\alpha\,\beta}$ can be chosen in
such a way that  $\sum_i\Lambda_i=1$, which implies of course five
linear equations for the  $n_{\alpha\,\beta}$'s. Their solution yields
\begin{align}
n_{00}&=\textstyle~\frac{-2}{(d-4+2\ell)(d-2-2\Delta)(d-3+\ell-\Delta)}\,,\nonumber\\
n_{20}&=\textstyle~\frac{2}{(d-4+2\ell)(d-2-2\Delta)(\ell+\Delta-1)}\,,\nonumber\\
n_{11}&=\textstyle~\frac{-8a\,b}{(d-4+\ell-\Delta)(d-2+\ell-\Delta)(\ell+\Delta-2)(\ell+\Delta)}\,,\\
n_{02}&=\textstyle~\frac{-2}{(d-4+2\ell)(d-2-2\Delta)(d+\ell-1)}\,,\nonumber\\
n_{22}&=\textstyle~\frac{2}{(d-4+2\ell)(d-2-2\Delta)(d-3+\ell-\Delta)}\,.\nonumber
\end{align}

Taking advantage of this normalization we finally obtain the KRT identity
\beq
 g^{(d-2)}_{\Delta,\ell}\equiv\sum_{i=1}^5\Lambda_i\,g^{(d-2)}_{\Delta.\ell}=\sum_{\alpha,\beta\in\{0,1,2\},\,\alpha+\beta\,\,{\rm
     even}}
k_{\alpha\,\beta}\, g^{(d)}_{\Delta+\alpha,\ell-\beta}~.
\eeq
 The $k_{\alpha\,\beta}$'s depend of course on the normalization of the
 conformal blocks. In accordance with \cite{Kaviraj:2019tbg} we have

\begin{align}
                k_{20}&=\textstyle -\frac{(\Delta -1) \Delta  \left(\Delta -\Delta
_{12}+\ell\right) \left(\Delta +\Delta _{12}+\ell\right) \left(\Delta -\Delta
_{34}+\ell\right)
                        \left(\Delta +\Delta _{34}+\ell\right)}{4 (d-2 \Delta -4) (d-2 \Delta -2) (\Delta
+\ell-1) (\Delta +\ell)^2 (\Delta +\ell+1)} \, ,\nonumber \\%[5pt]
                k_{11}&=\textstyle-\frac{(\Delta -1) \Delta _{12} \Delta _{34} \ell}{(\Delta
+\ell-2) (\Delta +\ell) (d-\Delta +\ell-4) (d-\Delta +\ell-2)} \,
,        \label{coefficients}\\%[5pt]
                k_{02}&=\textstyle-\frac{(\ell-1) \ell}{(d+2 \ell-6) (d+2 \ell-4)} ,\,~~~k_{00}=1
,\nonumber\\%[5pt]
                k_{22}&=\textstyle
                \frac{(\Delta -1) \Delta  (\ell-1) \ell \left(d-\Delta -\Delta _{12}+\ell-4\right)
\left(d-\Delta +\Delta _{12}+\ell-4\right) \left(d-\Delta
                        -\Delta _{34}+\ell-4\right) \left(d-\Delta +\Delta _{34}+\ell-4\right)}{4 (d-2
\Delta -4) (d-2 \Delta -2) (d+2 \ell-6) (d+2 \ell-4) (d-\Delta +\ell-5)
                        (d-\Delta +\ell-4)^2 (d-\Delta +\ell-3)} \, .\nonumber
        \end{align}

It is worth noting that the KRT identity simplify considerably  for scalar
conformal blocks (i.e. $\ell=0$),
where only  two terms remain. This simplification  can be traced back
to Eq.(\ref{f3}), which shows that the  operator $Y_3$ reduces the spin
$\ell$ by one. Since there are no conformal blocks with
negative spin, $Y_3$ annihilates any scalar conformal block. Furthermore,
 $Y_4$ and $Y_5$ are proportional to $Y_3$ or its derivatives, 
so they also annihilate this kind of conformal block. Consequently, in the 
scalar case, the iterative construction of the $Y_i$'s terminates at
$i=2$. The two remaining $\Lambda$ operators become in this case
\beq
\Lambda_{00}^{(\ell=0)}
  =\frac{Y_2+d-2-\Delta}{d-2-2\Delta}\,,\,
\Lambda_{20}^{(\ell=0)}=\frac {-Y_2-\Delta}{d-2-2\Delta}\,, \Lambda^{(\ell=0)}_{00}+\Lambda^{(\ell=0)}_{20}=1\,.
\eeq

Similar, simpler dimensional uplift identities for two-point and
three-point functions of CFT with boundary \cite{Zhou:2020ptb,Chen:2023oax} and  for
two-point functions in real projective space \cite{Giombi:2020xah}  have been found.
All of them can be reformulated in terms of differential operators
transforming a conformal block in dimension $d-2$ into a conformal block
in dimension $d$. For instance, in the case of CFT with boundaries, the
boundary conformal blocks $f_{bdy}^{(d)}[\Delta,\xi]$ are
eigenfunctions of the following quadratic Casimir operator
\beq
\cC[d]\,=\,-\xi(\xi+1)\p_\xi^2-d(\xi+\frac12)\p_\xi~,~\cC[d]\,f^{(d)}_{bdy}[\Delta,\xi]=\,\Delta\,(d-1-\Delta)\,f_{bdy}^{(d)}[\Delta,\xi]
\eeq
This operator acts linearly on $\{y_1=1,y_2=-(1+2\xi)\,\p_\xi\}$,
namely
\begin{align}
&\cC[d]\,y_1=\,y_1\,\cC[d-2]+y_2\\
&\cC[d]\,y_2=\,y_2\,(\cC[d-2]+2(d-3))-4y_1\,\cC[d-2]\,.
\end{align}
By applying the same approach as before we build two differential
operators, $\lambda_0$ and $\lambda_2$,  
\beq
\lambda_0=\frac{y_2+2(\Delta-d+3)}{2(2\Delta -d+3)}\,,~~\lambda_2=\frac{-y_2+2\Delta}{2(2\Delta -d+3)}\,\,\,,~~~\lambda_0+\lambda_2=1\,\,\,.
\eeq
such that \footnote{Since it has been noticed \cite{Gliozzi:2015qsa} that the
  three-dimensional boundary conformal blocks  are expressible as
  elementary algebraic functions, the mere existence of $\lambda_0$ and
  $\lambda_2$ allows us to conclude that this property holds for any
  odd $d$.}
\beq
\lambda_0\,f_{bdy}^{((d-2)}[\Delta,\xi]=f_{bdy}^{(d)}[\Delta,\xi]\,~
~,\,\lambda_2\,f_{bdy}^{(d-2)}[\Delta,\xi]=k\,f_{bdy}^{(d)}[\Delta+2,\xi]\,\,,
\eeq
where $k$ is a normalization-dependent coefficient. If for $\xi\to\infty$
$f_{bdy}^{(d)}[\Delta,\xi]\simeq\,1/\xi^\Delta$, we
have $k=\frac{\Delta(\Delta+1)}{4(d-2\Delta-5)(d-2\Delta-3)}$.

\section{Conclusions}

We provided a straightforward algebraic proof of a remarkable identity discovered
some time ago by Kaviraj, Rychkov and Trevisani (KRT). This identity expresses a
conformal block contributing to the four-point function in $(d-2)$ dimensions as a
linear combination of five conformal blocks in $d$ dimensions. The key mechanism
involved the iterative construction of five differential operators that transform
linearly under the action of the quadratic Casimir. This, in turn, enabled us to
define five differential operators, denoted by $\Lambda_i$, which transform any
conformal block in $(d-2)$ dimensions into five conformal blocks in $d$ dimensions.
These operators can be normalized such that $\sum_i\,\Lambda_i=1$, from
which the KRT identity follows immediately. Similar, simpler identities have been
proposed in recent literature, all of which can be reformulated using
the present approach.
As an example, we explicitly worked out the case of boundary conformal blocks
contributing to the two-point functions of boundary conformal field theories.

\bibliographystyle{abe}

\begin{thebibliography}{20}

\bibitem{Grinstein:1976zz}
G.~Grinstein and A.~Luther,
``Application of the renormalization group to phase transitions in disordered
systems,''
Phys. Rev. B \textbf{13} (1976), 1329-1343
doi:10.1103/PhysRevB.13.1329
%74 citations counted in INSPIRE as of 06 Apr 2025

%\cite{Aharony:1976jx}
\bibitem{Aharony:1976jx}
A.~Aharony, Y.~Imry and S.~K.~Ma,
``Lowering of Dimensionality in Phase Transitions with Random Fields,''
Phys. Rev. Lett. \textbf{37} (1976), 1364-1367
doi:10.1103/PhysRevLett.37.1364
%110 citations counted in INSPIRE as of 06 Apr 2025

%\cite{Parisi:1979ka}
\bibitem{Parisi:1979ka}
G.~Parisi and N.~Sourlas,
``Random Magnetic Fields, Supersymmetry and Negative Dimensions,''
Phys. Rev. Lett. \textbf{43} (1979), 744
doi:10.1103/PhysRevLett.43.744
%529 citations counted in INSPIRE as of 06 Apr 2025



%\cite{Imbrie:1984ki}
\bibitem{Imbrie:1984ki}
J.~Z.~Imbrie,
``Lower Critical Dimension of the Random Field Ising Model,''
Phys. Rev. Lett. \textbf{53} (1984), 1747
doi:10.1103/PhysRevLett.53.1747
%50 citations counted in INSPIRE as of 06 Apr 2025


%\cite{Brezin:1998fb}
\bibitem{Brezin:1998fb}
E.~Brezin and C.~De Dominicis,
``New phenomena in the random field Ising model,''
EPL \textbf{44} (1998), 13-19
doi:10.1209/epl/i1998-00428-0
[arXiv:cond-mat/9804266 [cond-mat]].
%19 citations counted in INSPIRE as of 06 Apr 2025

%\cite{Tissier:2011mv}
\bibitem{Tissier:2011mv}
M.~Tissier and G.~Tarjus,
``Nonperturbative Functional Renormalization Group for Random Field Models. IV:
Supersymmetry and its spontaneous breaking,''
Phys. Rev. B \textbf{85} (2012), 104203
doi:10.1103/PhysRevB.85.104203
[arXiv:1110.5500 [cond-mat.stat-mech]].
%47 citations counted in INSPIRE as of 06 Apr 2025

%\cite{Baczyk:2013uda}
\bibitem{Baczyk:2013uda}
M.~Baczyk, G.~Tarjus, M.~Tissier and I.~Balog,
``Fixed points and their stability in the functional renormalization group of
random field models,''
J. Stat. Mech. \textbf{1406} (2014), P06010
doi:10.1088/1742-5468/2014/06/P06010
[arXiv:1312.6375 [cond-mat.dis-nn]].
%7 citations counted in INSPIRE as of 06 Apr 2025

\bibitem{Cardy:1985}
John L. Cardy,
''Nonperturbative aspects of supersymmetry in statistical mechanics,'' Physica D:
Nonlinear Phenomena
\textbf{15} (1985), 123-128,
ISSN 0167-2789,
https://doi.org/10.1016/0167-2789(85)90154-X.


%\cite{Fytas:2019zdk}
\bibitem{Fytas:2019zdk}
N.~G.~Fytas, V.~Mart\'\i{}n-Mayor, G.~Parisi, M.~Picco and N.~Sourlas,
%``Evidence for Supersymmetry in the Random-Field Ising Model at D=5,''
Phys. Rev. Lett. \textbf{122} (2019) no.24, 240603
doi:10.1103/PhysRevLett.122.240603
[arXiv:1901.08473 [cond-mat.stat-mech]].
%26 citations counted in INSPIRE as of 06 Apr 2025

%\cite{Kaviraj:2019tbg}%\cite{Kaviraj:2019tbg}
\bibitem{Kaviraj:2019tbg}
A.~Kaviraj, S.~Rychkov and E.~Trevisani,
``Random Field Ising Model and Parisi-Sourlas supersymmetry. Part I. Supersymmetric
CFT,''
JHEP \textbf{04} (2020), 090
doi:10.1007/JHEP04(2020)090
[arXiv:1912.01617 [hep-th]].
%36 citations counted in INSPIRE as of 06 Apr 2025

%\cite{Kaviraj:2020pwv}
\bibitem{Kaviraj:2020pwv}
A.~Kaviraj, S.~Rychkov and E.~Trevisani,
``Random field Ising model and Parisi-Sourlas supersymmetry. Part II.
Renormalization group,''
JHEP \textbf{03} (2021), 219
doi:10.1007/JHEP03(2021)219
[arXiv:2009.10087 [cond-mat.stat-mech]].
%25 citations counted in INSPIRE as of 06 Apr 2025




%\cite{Kaviraj:2021qii}
\bibitem{Kaviraj:2021qii}
A.~Kaviraj, S.~Rychkov and E.~Trevisani,
``Parisi-Sourlas Supersymmetry in Random Field Models,''
Phys. Rev. Lett. \textbf{129} (2022) no.4, 045701
doi:10.1103/PhysRevLett.129.045701
[arXiv:2112.06942 [cond-mat.stat-mech]].
%19 citations counted in INSPIRE as of 06 Apr 2025



%\cite{Wiese:2021qpk}
\bibitem{Wiese:2021qpk}
K.~J.~Wiese,
``Theory and experiments for disordered elastic manifolds, depinning, avalanches,
and sandpiles,''
Rept. Prog. Phys. \textbf{85} (2022) no.8, 086502
doi:10.1088/1361-6633/ac4648
[arXiv:2102.01215 [cond-mat.dis-nn]].
%21 citations counted in INSPIRE as of 06 Apr 2025


%\cite{Parisi:1980ia}
\bibitem{Parisi:1980ia}
G.~Parisi and N.~Sourlas,
``Critical Behavior of Branched Polymers and the Lee-Yang Edge Singularity,''
Phys. Rev. Lett. \textbf{46} (1981), 871
doi:10.1103/PhysRevLett.46.871
%70 citations counted in INSPIRE as of 06 Apr 2025


%\cite{Cardy:2001ci}
\bibitem{Cardy:2001ci}
J.~L.~Cardy,
``Exact scaling functions for selfavoiding loops and branched polymers,''
J. Phys. A \textbf{34} (2001), L665-L672
doi:10.1088/0305-4470/34/47/101
[arXiv:cond-mat/0107223 [cond-mat]].
%9 citations counted in INSPIRE as of 15 Apr 2025

\bibitem{Cardy:2003}
J.~L. ~Cardy, ``Lecture on branched polymers and dimensional reduction'',
(2003), cond-mat/0302495.

\bibitem{hsu2005}
 H.~-P.~ Hsu, W.~ Nadler and P.~ Grassberger, ``Statistics of lattice
animals'', Comput. Phys. Commun. \textbf{169} (2005), 114–116.

\bibitem{luther2011}
 S.~ Luther and S.~ Mertens, ``Counting lattice animals in high
dimensions'', J. Stat. Mech. 2011 (2011), P09026.


%\cite{Kaviraj:2022bvd}
\bibitem{Kaviraj:2022bvd}
A.~Kaviraj and E.~Trevisani,
``Random field \ensuremath{\phi}$^{3}$ model and Parisi-Sourlas supersymmetry,''
JHEP \textbf{08} (2022), 290
doi:10.1007/JHEP08(2022)290
[arXiv:2203.12629 [hep-th]].
%8 citations counted in INSPIRE as of 06 Apr 2025

%\cite{Cardy:2023lha}
\bibitem{Cardy:2023lha}
J.~L.~Cardy,
``The Yang-Lee Edge Singularity and Related Problems,''
[arXiv:2305.13288 [cond-mat.stat-mech]].
%5 citations counted in INSPIRE as of 15 Apr 2025

\bibitem{Brydges2003a} 
D.C.~ Brydges and J.Z.~ Imbrie, ``Branched polymers and dimensional
reduction'', Ann. Math. \textbf{158} (2003), 1019–1039.

\bibitem{Brydges2003b}
 D. C.~ Brydges and J. Z.~ Imbrie, ``Dimensional Reduction Formulas for Branched
Polymer Correlation Functions'', J. Stat. Phys. \textbf{110} (2003) 503-518.



%\cite{Trevisani:2024djr}
\bibitem{Trevisani:2024djr}
E.~Trevisani,
``The Parisi-Sourlas uplift and infinitely many solvable 4d CFTs,'' SciPost Phys.
\textbf{18} (2025) no.2, 056
doi:10.21468/SciPostPhys.18.2.056
[arXiv:2405.00771 [hep-th]].
%5 citations counted in INSPIRE as of 15 Apr 2025




%\cite{Dolan:2011dv}e{Trevisani:2024djr}
\bibitem{Dolan:2011dv}
F.~A.~Dolan and H.~Osborn,
``Conformal Partial Waves: Further Mathematical Results,''
[arXiv:1108.6194 [hep-th]].
%404 citations counted in INSPIRE as of 15 Apr 2025

%\cite{Hogervorst:2013kva}
\bibitem{Hogervorst:2013kva}
M.~Hogervorst, H.~Osborn and S.~Rychkov,
``Diagonal Limit for Conformal Blocks in $d$ Dimensions,''
JHEP \textbf{08} (2013), 014
doi:10.1007/JHEP08(2013)014
[arXiv:1305.1321 [hep-th]].
%96 citations counted in INSPIRE as of 15 Apr 2025

%\cite{Fitzpatrick:2011dm}{Gliozzi:2016ysv}{Gliozzi:2017hni}
\bibitem{Fitzpatrick:2011dm}
A.~L.~Fitzpatrick and J.~Kaplan,
``Unitarity and the Holographic S-Matrix,''
JHEP \textbf{10} (2012), 032
doi:10.1007/JHEP10(2012)032
[arXiv:1112.4845 [hep-th]].
%278 citations counted in INSPIRE as of 15 Apr 2025

%\cite{Gliozzi:2016ysv}
\bibitem{Gliozzi:2016ysv}
F.~Gliozzi, A.~Guerrieri, A.~C.~Petkou and C.~Wen,
``Generalized Wilson-Fisher Critical Points from the Conformal Operator Product
Expansion,''
Phys. Rev. Lett. \textbf{118} (2017) no.6, 061601
doi:10.1103/PhysRevLett.118.061601
[arXiv:1611.10344 [hep-th]].
%45 citations counted in INSPIRE as of 15 Apr 2025


%\cite{Gliozzi:2017hni}
\bibitem{Gliozzi:2017hni}
F.~Gliozzi, A.~L.~Guerrieri, A.~C.~Petkou and C.~Wen,
``The analytic structure of conformal blocks and the generalized
Wilson-Fisher fixed points,''
JHEP \textbf{04} (2017), 056
doi:10.1007/JHEP04(2017)056
[arXiv:1702.03938 [hep-th]].
%59 citations counted in INSPIRE as of 15 Apr 2025


%
\bibitem{Zhou:2020ptb}
X.~Zhou,
``How to Succeed at Witten Diagram Recursions without Really Trying,'' JHEP
\textbf{08} (2020), 077
doi:10.1007/JHEP08(2020)077
[arXiv:2005.03031 [hep-th]].
%24 citations counted in INSPIRE as of 15 Apr 2025



%\cite{Chen:2023oax}
\bibitem{Chen:2023oax}
J.~Chen and X.~Zhou,
``Aspects of higher-point functions in BCFT$_{d}$,''
JHEP \textbf{09} (2023), 204
doi:10.1007/JHEP09(2023)204
[arXiv:2304.11799 [hep-th]].
%7 citations counted in INSPIRE as of 15 Apr 2025


%\cite{Giombi:2020xah}
\bibitem{Giombi:2020xah}
S.~Giombi, H.~Khanchandani and X.~Zhou,
``Aspects of CFTs on Real Projective Space,''
J. Phys. A \textbf{54} (2021) no.2, 024003
doi:10.1088/1751-8121/abcf59
[arXiv:2009.03290 [hep-th]].
%30 citations counted in INSPIRE as of 15 Apr 2025

%\cite{Gliozzi:2015qsa}
\bibitem{Gliozzi:2015qsa}
F.~Gliozzi, P.~Liendo, M.~Meineri and A.~Rago,
``Boundary and Interface CFTs from the Conformal Bootstrap,''
JHEP \textbf{05} (2015), 036
[erratum: JHEP \textbf{12} (2021), 093]
doi:10.1007/JHEP05(2015)036
[arXiv:1502.07217 [hep-th]].
%192 citations counted in INSPIRE as of 16 Apr 2025



\end{thebibliography}

\end{document}